\documentclass[prd, twocolumn, nofootinbib, floatfix]{revtex4-1}

\usepackage{amsmath}
\usepackage{graphicx}
\usepackage{dcolumn}
\usepackage{bm}
\usepackage{epsfig}
\usepackage{amssymb,latexsym,mathrsfs}
\usepackage{graphicx}
\usepackage{color}
\usepackage{hyperref}
\usepackage{float}
\usepackage{diagbox}

\hypersetup{
    colorlinks=true,
    linkcolor=red,
    citecolor=blue,
} 

\newcommand{\be}{\begin{equation}}
\newcommand{\ee}{\end{equation}}
\newcommand{\bs}{\begin{split}} 
\newcommand{\bea}{\begin{eqnarray}}
\newcommand{\eea}{\end{eqnarray}}

\newcommand{\al}{\alpha} 
 
\newcommand{\kp}{\kappa}

\newcommand{\event}{Even$t^{-1}$}

\begin{document}

\title{Finite Energy but Infinite Entropy Production from Moving Mirrors}
\author{Michael R.R. Good${}^{1,2}$}
\author{Eric V.\ Linder${}^{2,3}$} 
\affiliation{${}^1$Physics Department, Nazarbayev University, Astana, 
Kazakhstan\\
${}^2$Energetic Cosmos Laboratory, Nazarbayev University, Astana, 
Kazakhstan\\ 
${}^3$Berkeley Center for Cosmological Physics \& Berkeley Lab, 
University of California, Berkeley, CA 94720, USA} 

\begin{abstract} 
Accelerating mirrors provide a simple conceptual laboratory for studying particle 
production and the relation between trajectory and particle, energy, and entropy fluxes. 
We focus on the relation between energy and entropy, studying some special cases with finite total energy 
but infinite integrated entropy (though the entropy flux may be finite at any particular moment). 
We present a new asymptotically static moving mirror trajectory with solvable beta Bogolyubov coefficients, total energy and fully relativistic particle count.  The integrated entropy diverges despite finite global radiative particle and energy emission.  Another class of models includes exponentially accelerated mirrors 
in proper time; one of its unexpected behaviors is finite energy emission but divergent entropy. 
We compare mirrors exponentially accelerated in other coordinates as well, showing their close relation 
and an interesting duality property. 
\end{abstract} 

\date{\today} 

\maketitle


\section{Introduction}

Particle production from vacua in spacetime, e.g. \cite{Parker:1968mv,Hawking:1974sw,Unruh:1976db}, is a fascinating aspect of quantum field theory, connecting dynamics, 
energy flux, and information. One of the simplest systems for investigating these concepts is the 
accelerating mirror in 1+1 dimensional spacetime with scalar particle production \cite{DeWitt:1975ys,Davies:1976hi,Davies:1977yv}. Since few analytic 
solutions are known, new ones can give useful insights into the relations between these quantities. 

Of particular interest is entropy and its connection to information \cite{Chen:2017lum,Good:2016horizonless,Holzhey:1994we,Hwang:2017yxp,Hayden:2007cs}. Accelerating mirrors can generate 
analog black hole solutions, e.g. \cite{Good:2016MRB,Good:2016LECOSPA,MG14one,MG14two}, allowing exploration of the formation, emission, and perseverance or decay 
of black holes, and the associated information content of the particles produced. We focus here on the 
special situations where not only energy flux at a given moment, but the total energy emitted, is finite, 
yet the integrated entropy is infinite (and even the entropy flux may diverge). 

This has a twofold purpose. First, few solutions with finite total energy, and fewer still with analytically 
calculable energy and total finite particle production, are known \cite{walkerdavies,Good:2013lca,paper1}, so new solutions can help reveal the similarities 
and differences between their innate properties. Second, examining the relation between the energy and 
particle flux \cite{Walker:1984vj}, and their integrated quantities, and the entropy flux and integrated entropy, at the level 
of major disparity such as one being finite and the other infinite, offers a ``stress test'' to 
simple assumptions about their connection 
\cite{Hotta:2015yla}. Ideally these steps can eventually provide some further clarity 
on the fundamental relation between particle production and information. 

In Section~\ref{sec:betak} we present a new analytically solvable, asymptotically static mirror with 
interesting physical properties, and study its particle, energy, and entropy production. We contrast this 
with finite radiation, asymptotically null and drifting mirrors based on exponential acceleration in 
Section~\ref{sec:expprop}. Section~\ref{sec:sumtable} summarizes the diversity of behaviors in entropy 
despite the similarity of characteristics in energy flux or total energy. We discuss some future prospects 
and conclude in Section~\ref{sec:concl}.

\section{Finite Radiation, Asymptotically Static Solution} \label{sec:betak} 

Asymptotically static mirrors are of particular interest because they should have finite 
total energy production. However, this class is somewhat difficult to explore because the literature 
has only three solved\footnote{Here `solved', means only those trajectories where the beta 
Bogolyubov coefficients are analytically known, allowing calculation of particle flux and total number 
of particles.} asymptotic static mirrors: the Walker-Davies (1982) \cite{walkerdavies}, Arctx (2013) 
\cite{Good:2013lca}, and the self-dual solutions (2017) \cite{paper1}. 

Here we present a new solution that is also asymptotically static and with finite radiation, but simpler, more tractable, and more general 
in some respects 
than the first two previously known solutions.  It is also has significantly different, time-asymmetric dynamics than the aforementioned recent third solution. 

Note there is, as yet, no known exactly one-to-one analytically demonstrated correspondence\footnote{The correspondence exists for an asymptotically null trajectory \cite{Good:2016MRB,Good:2016LECOSPA,MG14one,MG14two}. Explicit derivations of the collapsing shell stress tensor in different vacuums can be found in \cite{Juarez-Aubry:2018ofz}. } to black hole particle production for asymptotically static trajectories which solve the soft particle production problem (e.g.\ \cite{Hawking:2016msc}) and represent complete evaporation with no left-over remnant \cite{Wilczek:1993jn}, so it is worthwhile exploring such cases 
further.

\subsection{New Mirror: betaK} 

Asymptotically static mirrors have useful physical properties so it is worthwhile attempting new solutions \cite{paper1}. Notably, they have total finite particle emission, avoiding the soft particle \cite{Hawking:2016msc} production problem.
We have found a new solution we call betaK, due to its exactly solvable beta Bogolyubov coefficients that take 
the form of a modified Bessel function. In addition it has some other advantages over the previous studied motions. The betaK trajectory is given by 
\bea 
z(t) &=& -\frac{v}{\kp} \sinh^{-1}(\kp t) = -\frac{v}{\kappa} \ln \left(\sqrt{\kp^2t^2+1}+\kp t\right)  \label{CS1z(t)}\\ 
\dot z &=& \frac{-v}{\sqrt{\kp^2 t^2+1}} \ , 
\eea 
where $z$ is the spatial coordinate, $t$ the time coordinate, $\kappa$ a scaling parameter that in the 
black hole case would be related to the surface gravity\footnote{The parameter $\kappa$ is \textit{not\/} the acceleration of the moving mirror, as it is for the uniformly accelerated observer in the thermal Unruh effect \cite{Unruh:1976db}. The thermal moving mirror \cite{Carlitz:1986nh} has proper acceleration $\alpha(\tau) = \tau^{-1}$, independent of $\kappa$-scale \cite{paper2}.} $\kappa \equiv (4M)^{-1}$, and $v$ is the maximum velocity of the 
mirror, occurring at $t=0$. One can readily see that at asymptotically large times, past and future, the 
mirror becomes asymptotically static, $\dot z\to0$. Moreover the velocity is time symmetric (and the 
trajectory is time asymmetric). 

The advantages include the ability to numerically solve for $N(v)$ particle count, in  contrast to Arctx's non-functional particle count \cite{Good:2013lca}. In addition, $z(t)$ is manifestly invertible, in contrast to the Walker-Davies mirror trajectory \cite{walkerdavies}, $t(z)$, which is not transcendentally invertible for $z(t)$. 
Moreover, this trajectory is found to be much more numerically tractable for all its interesting quantities than either Walker-Davies or Arctx.

\subsubsection{Trajectory} 

\begin{figure}[h]
\centering 
\rotatebox{90}{\includegraphics[width=3.2in]{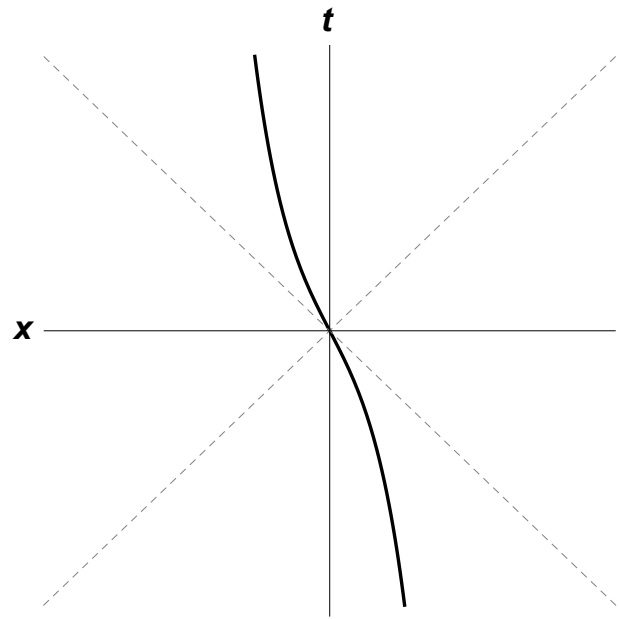}} 
\caption{The trajectory for betaK is asymmetric in time, and asymptotically static with finite energy and finite particle count. Here the maximum speed $v = 1/2$ and $\kappa = 1$.} 
\label{fig:CS1spacetime} 
\end{figure}

\begin{figure}[h]
\centering 
\includegraphics[width=3.2in]{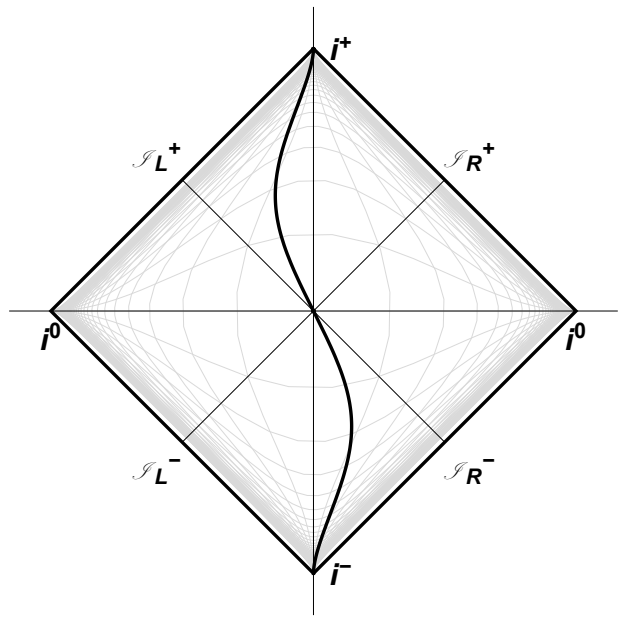} 
\caption{The trajectory for betaK, as in Fig.~\ref{fig:CS1spacetime}, but plotted in a conformal diagram.} 
\label{fig:CS1conformal} 
\end{figure}

The spacetime diagrams for the trajectory Eq.~(\ref{CS1z(t)}) are illustrated in 
Figure~\ref{fig:CS1spacetime} with a standard spacetime diagram and Figure~\ref{fig:CS1conformal} with a 
conformal or Penrose diagram.  The symmetries and asymptotically static character are reasonably evident in both (e.g.\ the mirror approaches time-like future infinity, $i^+$, along the vertical axis). 

This immediately classifies the dynamics of this solution with those `future and past asymptotically static trajectories' (see \cite{walkerdavies,Good:2013lca,paper1}). This motion is distinct from the typically infinite energy producing, late-time thermal trajectories of the `asymptotically null' solutions (see the black mirror \cite{Good:2016MRB,Good:2016LECOSPA,MG14one,MG14two} or the thermal mirror \cite{Carlitz:1986nh,paper2,Good:2012cp}).  Moreover, it exhibits regularizing behavior distinct even from
the asymptotically inertial, soft particle producing trajectories of asymptotically drifting solutions (see e.g. \cite{Good:2015nja,Good:2018GTC}).

\subsubsection{Energy Flux} 

The energy flux of particles produced is related to the trajectory by \cite{Davies:1977yv},
\be 
F(t)=-\frac{1}{12\pi}\frac{\dddot z (1-\dot z^2)+3\dot z \ddot z^2}{(1-\dot z)^4 
(1+\dot z)^2} \ . \label{eq:flux} 
\ee 
Computation of the flux emitted to the right of the mirror (by convention) from the stress tensor observed by measurement at future-null infinity, $\mathscr{I}^{+}_R$, is straightforward, 
\be 
F(t) = \frac{\kappa^2}{12\pi}\frac{v\sqrt{\kp^2t^2+1}  \left(2 \kp^2t^2+v^2-1\right)}{\left(v-\sqrt{\kp^2t^2+1}\right)^2 \left(v+\sqrt{\kp^2t^2+1}\right)^4}\ . 
\ee 
This has a central valley of negative energy flux, as is a well-known \cite{Ford:1990ae} feature of moving mirrors. It has positive energy emission approaching  
the $t\to \pm \infty$ asymptotes, with total positive energy as seen in the next subsection.

\subsubsection{Total Energy}

Due to the asymptotically static character, the total energy is finite.  The total energy emitted from the right side of the accelerating mirror is analytically calculable,
\be 
E_R = \frac{\kappa}{96\pi}\frac{\gamma^2}{v^2} \left[\gamma  \left(6-8 v^2\right) \sin ^{-1}v+\pi  \gamma  v^4+4 v^3-6v\right], 
\ee 
where $\gamma \equiv (1-v^2)^{-1/2}$ is the Lorentz factor (again, $v$ is the maximum speed of the mirror).
Accounting for both sides, the total energy $E_T= E_R + E_L$ takes the remarkably simple expression  
\be E_T = \frac{\kappa \gamma^3 v^2}{48}\ , \label{eq:etotal} \ee
demonstrating immediately three 
physical results: 1) zero maximum speed gives zero energy (no particle production), 2) the total 
energy is positive, and 3) 
an arbitrarily fast mirror, $v \to 1$, gives divergence of energy production due to the Lorentz factor.

\subsubsection{Particle Flux} 

Quite unusually, the beta Bogolyubov coefficients describing particle emission can be solved for, with 
the fairly simple result 
\be \beta_R(\omega,\omega') = -\frac{2 v \sqrt{ \omega \omega'}}{\pi  \omega_p} e^{-\frac{\pi}{2}  v \omega_n} K_{i v \omega_n}(\omega_p)\ ,\ee
where $K_n(z)$ is a modified Bessel function of the second kind, 
$\omega_p \equiv \omega+\omega'$ and $\omega_n \equiv \omega - \omega'$, $\omega'$ and $\omega$ are 
the in and outgoing mode frequencies respectively, and $\kappa = 1$ for convenience. 

The particle spectrum per mode per mode (modulus squared) is
\be |\beta_R|^2 = \frac{4 v^2 \omega \omega'}{\pi ^2 \omega_p^2} e^{-\pi  v \omega_n}| K_{i v \omega_n}(\omega_p)|^2\ ,\label{rightspec}\ee
and accounting for both sides, $|\beta_T|^2 = |\beta_R|^2 + |\beta_L|^2$, gives 
\be |\beta_T|^2 = \frac{8 v^2  \omega  \omega' }{\pi ^2 \omega_p^2}\cosh (\pi  v \omega_n) |K_{i v \omega_n}(\omega_p)|^2\ . \ee
As a crosscheck, the total energy can be retrieved using the particles through globally summing quanta, 
\be E_T = \int_0^\infty \int_0^\infty  \omega \cdot |\beta_T|^2 \; d \omega \; d \omega'\ , \ee
which gives, reinstating the scale $\kappa$,
\be E_T = \frac{\kappa \gamma^3 v^2}{48}\ , \label{eq:etotalbeta} \ee 
demonstrating consistency of the solution with Eq.~(\ref{eq:etotal}).

\subsubsection{Particle Spectrum}
\begin{figure}[h]
\centering
\includegraphics[width=3.2in]{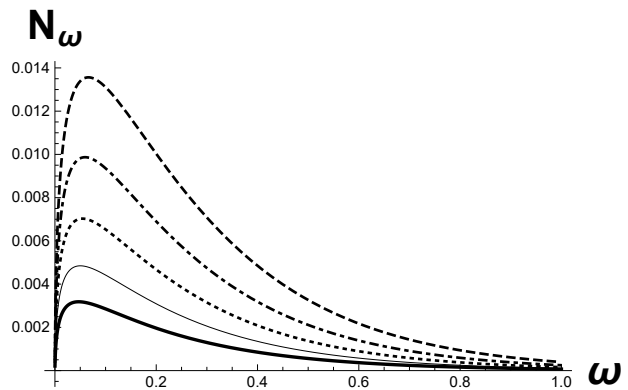} 
\caption{The spectrum, $N_\omega$, or particle count per mode detected at the right Cauchy surface, of the 
asymptotically static time-asymmetric mirror betaK, is plotted vs frequency 
in units of $\kappa$.  The different lines correspond to different maximum speed trajectories: $v =  0.5, 0.6, 0.7, 0.8, 0.9$ from bottom to top, for thick solid, thin solid, dotted, dot-dashed, and dashed, 
respectively. This asymptotically static solution has no infrared divergence (soft particles) suffered by mirrors that are asymptotically drifting.
\label{fig:spectrum}} 
\end{figure}

The spectrum, $N_\omega$, or particle count per mode, detected at the right Cauchy surface  $\mathscr{I}^{+}_R$ is found by inserting Eq.~(\ref{rightspec}) into
\be N_\omega = \int_0^\infty  |\beta_R|^2 \; d \omega'\ . \ee
Figure~\ref{fig:spectrum} illustrates the results for different maximum mirror speeds.

\subsubsection{Total Particles}

\begin{figure}[h]
\centering 
\includegraphics[width=3.2in]{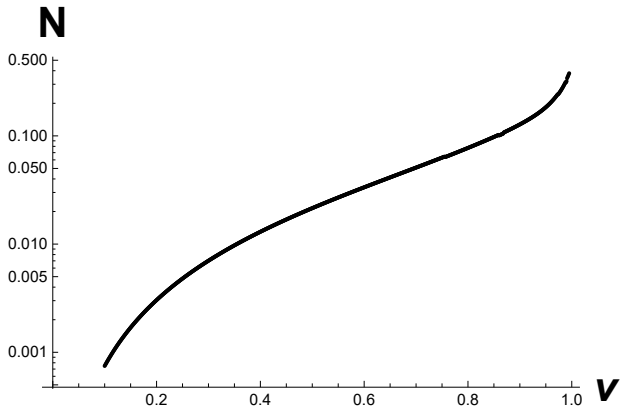} 
\caption{The time-asymmetric mirror betaK, being asymptotically static, has finite particle count. 
The total particle emission count is plotted for trajectories with relativistic maximum speeds $v$. 
}
\label{fig:particleCS1} 
\end{figure}

The particle count $N(v)$ from the mirror with maximum speed $v$ comes from summing over the spectrum, 
\be N(v) = \int_0^\infty \int_0^\infty  |\beta_T|^2 \; d \omega \; d \omega' .\label{Nofv} \ee
Unusually, this turns out to be finite and numerically tractable for any choice of maximum speed, 
$0\leq v < 1$. The sharp cutoff for soft and hard massless scalar particles, i.e.\ at low and high $\omega$, plays a role 
in this; the total number of particles from accelerating mirrors commonly is infinite even when the total 
energy is finite (even for asymptotically inertial-drifting mirrors). Figure~\ref{fig:particleCS1} shows the behavior of $N(v)$, Eq.~(\ref{Nofv}). The number is small, 
$N(v) < 1$ (recall particle number is dimensionless and so the result is independent of the dimensional 
scale-parameter $\kappa$), and increases monotonically with the chosen maximum speed $v$.

\subsubsection{Entropy Flux} 

It is noteworthy that emission of von Neumann entanglement entropy does not change sign for this solution, Eq.~(\ref{CS1z(t)}).  The entropy flux (see e.g. \cite{Chen:2017lum,Good:2016horizonless} and references therein),
\be S(t) = -\frac{1}{6} \tanh^{-1} \dot{z}(t)\ ,\ee 
radiated to the right (left) is always positive (negative), with 
\be 
S(t)_{R,L} = \pm \frac{1}{6}\tanh^{-1}\left(\frac{v}{\sqrt{\kp^2 t^2+1}}\right) \ . \label{CS1S(t)} 
\ee 

\begin{figure}[h]
\centering 
\includegraphics[width=3.2in]{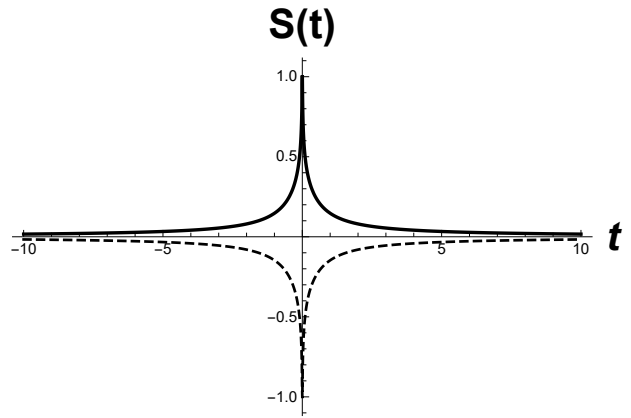} 
\caption{The time-asymmetric mirror betaK from Eq.~(\ref{CS1z(t)}), being asymptotically static with finite energy and finite particle count, might be expected to have finite integrated entropy, however this is not the case. The entropy flux is plotted as the solid line (dashed line) and is always positive (negative) as emitted from the right (left) side of the mirror. Here the flux $S(t)$ from Eq.~(\ref{CS1S(t)}) is shown for the choice of maximum speed $v = 0.99999$ and $\kappa = 1$. See the text for discussion of the integrated entropy integral and its divergence.}
\label{fig:entropyCS1} 
\end{figure}

Figure \ref{fig:entropyCS1} illustrates this entropy flux. Note that in the far past 
and future, where $\kp |t|\gg 1$, the entropy flux $S(t)\sim 1/|t|$. This property 
will be important for the next subsection, and the general comparison of energy 
characteristics to entropy characteristics.

\subsubsection{Integrated Entropy} \label{sec:cs1intent} 

Despite the finite energy, and finite particle emission, and the preservation of unitarity, the integrated amount of entropy diverges. This is given by  
\be S_I = \int_{-\infty}^\infty S(t)\; (1-\dot z)\,dt \ , \ee
and since $\dot z\to0$ at large times, we see the divergence arises from $S(t)$ 
itself. 

We can understand the divergence mathematically by noting that in the previous subsection we saw that the entropy flux only dies as $1/t$ for large 
$|t|$, and hence the integrated entropy has a logarithmic 
divergence. More generally, for an asymptotically static mirror (where $\dot z\to0$), if $\dot z\sim t^{-n}$ at late times with $n>0$ then the proper acceleration $\alpha\equiv \gamma^3\ddot z\sim t^{-n-1}$ and the rapidity $\eta=\tanh^{-1}\dot z$ and entropy flux $S(t)$ have late time 
contributions going as $t^{-n}$. Note that unitarity is preserved for $n>0$. Then the integrated entropy $S_I\sim t^{-n+1}$ and hence 
diverges if $n<1$ (for $n=1$, $S_I$ diverges logarithmically -- this is precisely the 
betaK behavior). We can also calculate that the energy flux will die off as $t^{-n-2}$ 
and so the total energy gets a late time contribution going as $t^{-n-1}$. 
Thus betaK (which has $n=1$) represents the ``boundary'' case between the mirror giving both finite energy and 
finite integrated entropy and the one producing finite energy but infinite integrated entropy. 

The physical interpretation of this is less clear. The mirror has a finite particle count and energy. It may be that it has an infinite number of particle 
states with infinitesimal mean occupation, so $N$ is finite but since $S_I$ counts the 
number of states it is divergent.

\subsection{Comparison to Related Finite Energy Mirrors} 

We identified above the asymptotic behavior of $\dot z\sim t^{-1}$ 
as a key ingredient for finite energy but infinite integrated entropy. Let us 
explore this further by comparing the betaK case to two other examples 
with the same approach to the asymptotic static limit. 

One is the self dual solution of \cite{paper1}. This has 
\be 
\dot z=\frac{2v\kp t}{\kp^2 t^2+1} \ , 
\ee 
which indeed has the same behaviors of $\dot z\sim t^{-1}$, $\alpha\sim t^{-2}$, 
$F\sim t^{-3}$, and $S(t)\sim t^{-1}$. However, because $\dot z$ and 
hence $\eta$ and $S(t)$ are odd in time, the logarithmic divergence in 
the integral for the total entropy cancels and the total integrated entropy was 
found in \cite{paper1} to be finite. So time asymmetry vs symmetry plays an important role. 

For direct comparison to betaK we therefore we need to study an even 
function $\dot z$ that still asymptotes as $\dot z\sim |t|^{-1}$. 
The simplest instance of this after betaK is a model we call \event,  
\bea 
\dot z&=&v\sqrt{\frac{27}{4}}\,\frac{\kp ^2 t^2}{(\kp^2 t^2+1)^{3/2}} \label{eq:sq32} \\ 
z&=&\frac{v}{\kp}\sqrt{\frac{27}{4}}\,\left[\frac{-\kp t}{\sqrt{\kp^2 t^2+1}} +\ln\left(\kp t+\sqrt{\kp^2 t^2+1}\right)\right] \,. 
\eea 
This has maximum velocity $v$ at $\kp^2 t^2=2$, and is asymptotically 
static with $\dot z\sim t^{-1}$. Again, the total energy is finite but, since $\dot z$ and hence $S(t)$ are even, 
the integrated entropy is again infinite. 

\begin{figure}[h]
\centering 
{\includegraphics[width=3.2in]{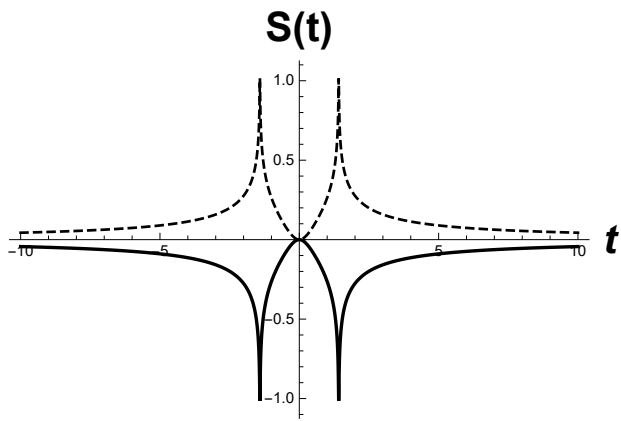}} 
\caption{The entropy for the \event\ mirror with trajectory Eq.~(\ref{eq:sq32}).  The trajectory is asymptotically static with finite energy but divergent integrated entropy, similar to betaK. The solid (dashed) line is the entropy to the right (left).   }
\label{fig:entropy_spikes} 
\end{figure}

Figure~\ref{fig:entropy_spikes} shows the entropy flux 
as a function of time, which can be compared to 
Fig.~\ref{fig:entropyCS1}. Both die off as $1/t$ for 
large times. While betaK has its maximum velocity for 
$t=0$, and hence a spike in entropy flux there, 
the \event\ model has maximum velocity and entropy spikes 
at $\kp^2 t^2=2$.

\section{Finite Radiation, Asymptotically Null and Drifting Dynamics} \label{sec:expprop} 

As a counterpoint to the previous section on asymptotically static mirrors, 
their finite energy, asymptotically vanishing entropy flux, and infinite 
integrated entropy, we consider an asymptotically null and then a drifting mirror. 

Asymptotically null mirrors have no guarantee of finite energy production 
but we develop a new solution that does, and has interesting thermodynamic 
properties, allowing us to study further the relation between energy and 
entropy. The solution employs exponential acceleration in proper time, and can be 
viewed as a new mirror in the series 
\begin{itemize}
	\item $\alpha(u) \sim e^u$ for Carlitz-Willey \cite{paper2,Carlitz:1986nh, Good:2012cp} 
    \item $\alpha(t) \sim e^t$ for Hotta-Shino-Yoshimura \cite{Good:2013lca, paper2, Hotta:1994ha} 
    \item $\alpha(x) \sim e^x$ for Davies-Fulling \cite{Davies:1977yv,paper2}  
\end{itemize} 

We investigate an exponentially accelerated trajectory for the mirror in proper time, 
\be \alpha(\tau) = - \kp e^{\kp \tau}\ ,\label{exptau} \ee
where the negative sign is by convention to send the mirror accelerating to the left. 

Intuition suggests an infinite total energy, similar to the results of the aforementioned exponentially asymptotically accelerated null mirrors.  Surely, a system which feels an ever increasing acceleration will produce ever increasing energy? Interestingly, with this equation of motion Eq.~(\ref{exptau}), this expectation is dashed.

\subsubsection{Trajectory Dynamics}

\begin{figure}[h]
\centering 
\includegraphics[width=3.2in]{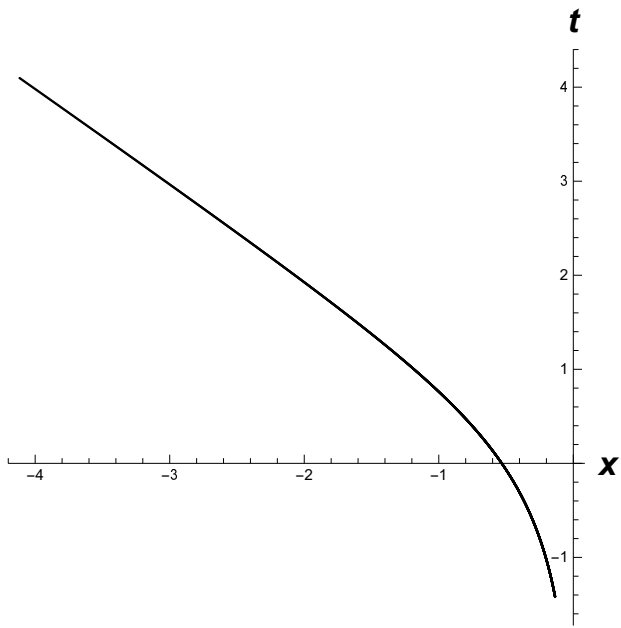} 
\caption{The exponential accelerating trajectory $z(\tau)$ from Eq.~(\ref{trajectory}) is plotted in a coordinate time $t$ spacetime diagram, with $\kappa = 1$. 
}
\label{fig:trajectory} 
\end{figure}

The dynamical trajectory functions are 
\bea 
\eta(\tau) &=& -e^{\kp \tau} \qquad\qquad\quad \gamma(\tau) = \cosh(e^{\kp \tau})\\ 
w(\tau) &=& - \sinh(e^{\kp \tau}) \qquad \textrm{v}(\tau) = - \tanh(e^{\kp \tau}) \\ 
z(\tau) &=& - \frac{1}{\kp} \textrm{Shi}(e^{\kp \tau}) \qquad t(\tau) = \frac{1}{\kappa}\textrm{Chi}(e^{\kp \tau})\ , \label{trajectory} 
\eea 
where the rapidity $\eta'(\tau)=\alpha(\tau)$, Lorentz factor $\gamma=\cosh\eta$, 
celerity (proper velocity) $w=dz/d\tau=\sinh\eta$, velocity $v=dz/dt=\tanh\eta$, and Shi (Chi) is the hyperbolic sine (cosine) integral. 
We plot the trajectory $z(t)$ function in Figure~\ref{fig:trajectory}. Note that 
for large $t$ (or $\tau$), $z\sim t$. The mirror velocity asymptotically approaches 
the speed of light, as to be expected.

\subsubsection{Energy Flux} 

\begin{figure}[h]
\centering 
\includegraphics[width=3.2in]{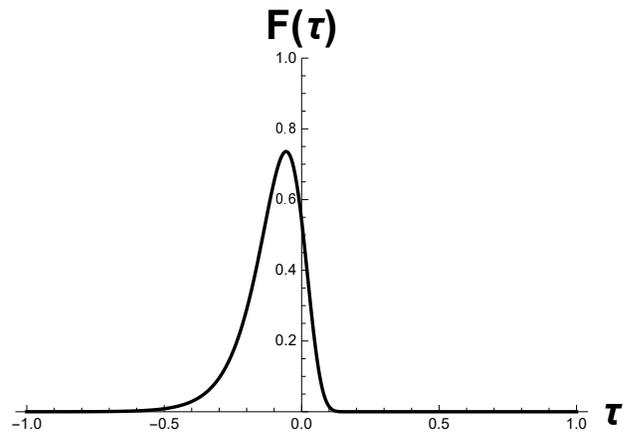} 
\caption{The energy flux to the right of the exponential accelerating mirror contains 
no negative energy flux and demonstrates terminal evaporation. Here $\kappa= \sqrt{48\pi}$, normalized so that thermal energy flux would be $F=1$.}  
\label{fig:flux} 
\end{figure}

To calculate the energy flux produced by this exponentially accelerating mirror, 
we use the energy flux relation $12\pi F(\tau) = - \alpha'(\tau) e^{2\eta(\tau)}$ \cite{paper2} to find 
\be F(\tau) = \frac{\kappa^2}{12\pi}  e^{\kappa  \tau -2 e^{\kappa  \tau }}\ . \label{energyflux}\ee 
The energy flux is plotted in Figure~\ref{fig:flux}. 
Note the emission is always positive -- there is no negative energy 
flux (NEF).  This is a particularly interesting case because unlike the no NEF solutions of Carlitz-Willey \cite{Carlitz:1986nh}, the black mirror \cite{Good:2016MRB}, or Hotta-Shino-Yoshimura (Arcx) \cite{Good:2013lca, paper2,Hotta:1994ha, BENITO} for example, in the far past and future the energy flux  asymptotes to zero, despite $\alpha(\tau) \sim e^{\tau}$. This indicates that the radiation process completely terminates (as far as energy evaporation is concerned); for a black hole analog this would correspond to evaporation with an asymptotically infinite Doppler-shifting remnant (a `super-remnant', if you will) consistent with the conservation of energy without backreaction \cite{Fabbri:2005mw}.

\subsubsection{Total Energy} 

Recall our criteria from Section~\ref{sec:cs1intent} for finite entropy flux and 
integrated entropy. For the asymptotically static mirror we wanted $\dot z$ to 
die off quicker than $1/t$, giving the acceleration dying quicker than 
$1/t^2$ and the flux dying quicker than $1/t^3$ in order to get both finite total 
energy and integrated entropy. Here, however, we have $\dot z$ going to a constant 
(the speed of light), acceleration exponentially increasing, but energy flux 
dying off rapidly. 

To continue the investigation, we calculate the total energy for an observer at $\mathscr{I}^+_R$, 
\be E = \int F(\tau)\; du\ , \ee
where $u=t-z$ is the null coordinate. This integral can be done and has a simple form. 
Since $du/d\tau=\cosh\eta-\sinh\eta=e^{-\eta}= e^{e^{\kp t}}$ then  
\be E = \frac{\kappa}{12\pi} \int_{-\infty}^{\infty}\kappa e^{\kappa  \tau -2 e^{\kappa  \tau }} e^{e^{\kappa\tau}} d\tau \ .\ee
The result of the integral is unity, and so the total energy emitted to the observer at $\mathscr{I}^+_R$ is finite, with 
\be E = \frac{\kappa}{12\pi} \ .\ee 
Thus the condition for finite energy seems to depend only the energy flux 
dying away sufficiently quickly, and not on the asymptotic behavior of individual 
trajectory dynamics quantities such as 
$\dot z$ or acceleration per se (though in combinations they do determine the flux). 

This result from the exponentially accelerating mirror has a drifting mirror counterpart \cite{Good:2016HUANG}, where the acceleration asymptotically approaches zero in the far future and the mirror can coast at the speed of light. There $E = \kappa/(96\pi)$ found in \cite{Good:2013lca}. 

The new mirror with exponential acceleration in proper time, Eq.~(\ref{exptau}), 
is the only one of the exponential forms mentioned to possess finite energy. It is surprising, without yet considering the entropy, that despite an ever increasing asymptotically infinite acceleration the system radiates a finite total energy.

\subsubsection{Entropy Flux and Integrated Entropy} 

The entropy flux, $S(\tau)$, is found from the rapidity $\eta(\tau) = -6 S(\tau)$ \cite{paper1, Good:2015nja}, so that
\be S(\tau) = \frac{1}{6}\,e^{\kp \tau}\ . \ee
It clearly diverges at late times, in stark contrast with the rapidly vanishing energy flux, Eq.~(\ref{energyflux}).  The integrated entropy is not saved by integration over $u$ at $\mathscr{I}^+_R$, as the integral 
\be S_I = \frac{1}{6} \int_{-\infty}^{\infty} e^{\kp \tau} e^{e^{\kp \tau}} d \tau  \ee 
also diverges.  This demonstrates a loosening between the information content and energy content carried by the radiation.  Despite the finite energy production, unitarity is lost because the entropy flux $S(\tau)$ does not asymptote to a constant, but diverges as $\tau \to +\infty$. 

Note that the entropy and the proper acceleration $\alpha = \kappa \eta$ simply scale together for the new exponential mirror, with 
\be \alpha = - 6 \kappa S\ . \ee
This is in contrast to the other exponential forms: Carlitz-Willey, Hotta-Shino-Yoshimura, and Davies-Fulling respectively have $\eta = -\kappa u/2$, $\eta =\kappa x$, and $\eta = -\kappa t$, so the entropy involves inverse hyperbolic trig functions of the acceleration.

\subsubsection{Exponential in $\tau$ Multiplicatively Shifted} 

We can use the technique of multiplicatively shifting the mirror trajectory, i.e.\ $\dot z \to v\dot z$, to regularize the infinite asymptotic acceleration \cite{paper1}. This takes the asymptotically null mirror 
to an asymptotically drifting one. As we just saw, if $\dot z=dz/dt=-\tanh(e^{\kappa\tau})$ then $\alpha(\tau)=-\kappa e^{\kappa\tau}$. This gave zero flux at late times but infinite entropy. But if we 
multiplicatively shift to 
\be 
\dot z= v\tanh(e^{\kp\tau}) \ , \label{eq:exptauv} 
\ee 
which has asymptotically constant velocity less than the speed of light, then the proper acceleration 
\be 
\al(\tau)=\frac{v\kp\, e^{\kp\tau}}{\cosh^2 (e^{\kp\tau})-v^2\sinh^2 (e^{\kp\tau})} \ , 
\ee 
which for any $v<1$ goes to zero for large $\tau$. So at large times, energy flux and acceleration goes to 0, and the rapidity $\eta$ and 
entropy flux $S(\tau)$ are finite, while the integrated entropy still diverges. 

As a broad principle relevant to the several cases discussed, for entropy there is 
a straightforward relation to the velocity $\dot z$, along the lines of the 
criterion in Section~\ref{sec:cs1intent}. Recall 
$S(\tau)=-(1/6)\eta=-(1/6)\tanh^{-1}\dot z$. When asymptotically 
$\dot z\to0$ then the entropy flux goes to zero, and if this proceeds 
quickly enough then the integrated entropy stays finite. For the exponentially 
accelerating mirror cases, $\dot z\to\,$const and so integrated entropy is infinite. In the drifting 
mirror subcase (i.e.\ exponential acceleration regularized to approach zero), with $v<1$, $S(\tau)$ stays finite while for 
the nonregularized $v=1$ case above we have $S(\tau)\sim\tanh^{-1}(1)\to\infty$.

\section{Summary of Results} \label{sec:sumtable} 

In Table~\ref{tab:entropy} we summarize the (mostly) new mirrors we have 
considered, showing how there can be a diversity of behaviors in 
entropy even with the same characteristics in energy flux or total energy.

\begin{table*} 
\begin{center} 
\begin{tabular}{l|c|c|c|c} 
Model & \ Energy Flux \ & \ Total Energy \ & \ Entropy Flux \ & \ Integrated Entropy \ \\ 
\hline  
\rule{0pt}{1.1\normalbaselineskip}betaK (Eq.~\ref{CS1z(t)}) & $\sim t^{-3}$ & $\kp v^2\gamma^2/48$ & $\sim t^{-1}$  & log divergent \\ 
Self-Dual \cite{paper1} & $\sim t^{-3}$ & \ $\kp v^2\gamma(\gamma^2+3)/48$ \ & $\sim t^{-1}$  & finite \\ 
\event\ (Eq.~\ref{eq:sq32}) & $\sim t^{-3}$ & finite & $\sim t^{-1}$ & log divergent \\ 
Exptau (Eq.~\ref{exptau}) & $\to0$ & $\kp/(12\pi)$ & diverges & infinite \\ 
Exptau(v) (Eq.~\ref{eq:exptauv})$\quad$ & $\to0$ & finite & finite & infinite \\ 
\end{tabular}
\end{center}
\caption{The energy and entropy properties are summarized for the models discussed in this article. 
The flux behaviors listed are those in the asymptotic future. All these models have finite 
total energy but differing entropy behaviors. Note the self dual solution 
avoids infinite total entropy through its self dual nature (symmetry in time). 
} 
\label{tab:entropy} 
\end{table*}

The first three mirrors are closely related in their properties, showing the 
``boundary'' case of energy flux dying off as $t^{-3}$ and entropy flux diminishing 
as $t^{-1}$. This leads to a logarithmic divergence in integrated entropy -- except 
for the self dual mirror which is saved by its time symmetry (i.e.\ self dual nature). 
If the flux fades more rapidly then the integrated entropy would be finite. The 
betaK and \event\ mirrors are new, with the betaK case of particular interest due to 
its solvable beta Bogolyubov coefficients and tractable particle production 
characteristics. 

The last two mirrors add to the exponential mirror family (which is completed in 
Appendix~\ref{sec:expv}), with Exptau being a new 
solution on a par with well known mirrors -- with the added attractions of having no 
negative energy flux, flux asymptoting to zero, and a particularly simple linear 
relation between proper acceleration and entropy. The Exptau(v) case is the drifting 
mirror sibling that regularizes the 
acceleration from infinity at large times to zero and keeps the entropy flux finite.

\section{Conclusions} \label{sec:concl} 

Particle production from accelerating mirrors by itself is a fascinating physics 
phenomenon, but its relation to entropy and information brings unexpected depths to 
the study of moving mirrors. We presented four new mirrors, comparing and contrasting 
their particle production, energy flux, entropy flux, and integrated entropy 
characteristics. 

Looking for a time-asymmetric finite particle creation solution, we found the betaK mirror which is only the fourth solved asymptotically static mirror, and has 
beta Bogolyubov coefficients of the form of a modified Bessel function. It has a simple 
expression for its finite total energy, and calculable finite total particle count, 
but infinite integrated entropy. This raises interesting questions regarding the exact 
relation between particle and energy production and information. A close relative is 
the \event\ mirror, slightly more complex and with different patterns of entropy flux 
though the same asymptotic behavior. We also compared these to the self dual mirror 
introduced in \cite{paper1}, which again has the same asymptotic energy and entropy 
flux behaviors but a finite integrated entropy due to its time symmetry. 

We presented general guidelines to the asymptotic behaviors in velocity, proper 
acceleration, energy flux, and entropy flux; in particular we identified a ``boundary'' 
behavior where when the velocity asymptotically vanishes more rapidly than 
$t^{-1}$, and hence the other three quantities asymptotically vanish more rapidly than $t^{-2}$, 
$t^{-3}$, and $t^{-1}$ respectively, the integrated entropy would remain finite. 

Moving from asymptotically static to asymptotically null and drifting mirrors, we studied 
Exptau, a new 
mirror in the exponential acceleration family (that includes the Davies-Fulling and 
Carlitz-Willey mirrors), this one exponential in proper time. It has no negative energy 
flux at any time, and the flux rapidly vanishes asymptotically, analogous to concluded 
evaporation (energy emission ends) of a black hole. Interestingly, the entropy is directly proportional to 
the acceleration, and becomes infinite. This seems to imply a disconnect, in this case 
at least and asymptotically, between information (presumably related to entropy) and 
the state of the black hole (which has evaporated). We also introduced a regularized 
variant, Exptau(v), that asymptotically drifts at less than the speed of light and 
has vanishing asymptotic acceleration. Its entropy flux remains finite, though its 
integrated entropy diverges. In the appendix we also completed the exponential family 
by investigating acceleration in advanced time $v$, and identifying interesting 
``duality''-like relations.

Considering future directions, as we have seen from investigating proper time exponential acceleration in Eq.~(\ref{exptau}), it could be useful to work with proper time in more general contexts, such as for the energy-entropy flux relations, which are easy to express in terms of both null time and proper time.  In terms of 
null time $u$, $\eta(u) = -6S(u)$ and we can write Eq.~(\ref{eq:flux}) as
\be F(u) = \frac{1}{2\pi}\,\left[6S'(u)^2 + S''(u)\right]\ , \ee 
and in terms of proper time we can use the relation $\eta(\tau) = - 6 S(\tau)$ to write 
\be F(\tau) = \frac{1}{2\pi} S''(\tau)\, e^{-12 S(\tau)}\ . \label{FtauS}\ee 
This result demonstrates a direct relationship between negative energy flux and entanglement entropy:  It is the sign of $S''(\tau)$ that determines the emission of NEF. 

The possible concavity of the entropy found here (see \cite{Holzhey:1994we} for a relation in terms of correlations) indicates the connection to the locally negative energy which emerges in the usual analysis of the static Casimir effect and of vacuum polarization near black hole horizons, yet in this moving mirror case, the negative energy is radiated.  

The simplicity of Eq.~(\ref{FtauS}) contains the deeper underlying symmetry of the model 
\cite{Fabbri:2005mw}, namely the M\"obius transformations of $SL(2,\mathbb{R})$,
\be p(u) \to \frac{ a p(u) + b }{c p(u) + d}, \quad ad-bc = 1, \ee
in the Schwarzian derivative  
\be -24\pi F(u) = \{p,u\} \equiv \frac{p'''}{p'} - \frac{3}{2}\left(\frac{p''}{p'}\right)^2, \ee
of the trajectory dynamics as encapsulated in the null-coordinate function, $p(u)$ (the $v$ position of the mirror as function of $u$).  We intend to explore this symmetry as connected to the SYK model (see e.g.\ \cite{Sarosi:2017ykf} and references therein) whose action also has this emergent conformal symmetry (in the IR, large $N$ limit), as a consequence of the two special properties: conformal flatness and conformal invariance \cite{Fulling:2018lez}.

\acknowledgments 

Funding from state-targeted program ``Center of Excellence for Fundamental and Applied Physics" (BR05236454) by the Ministry of Education and Science of the Republic of Kazakhstan is acknowledged. MG is funded by the ORAU FY2018-SGP-1-STMM Faculty Development Competitive Research Grant No. 090118FD5350 at Nazarbayev University. 
EL is supported in part by the Energetic Cosmos Laboratory and by 
the U.S.\ Department of Energy, Office of Science, Office of High Energy 
Physics, under Award DE-SC-0007867 and contract no.\ DE-AC02-05CH11231.

\appendix 

\section{Exponential acceleration in advanced time $v$} \label{sec:expv} 

In Section~\ref{sec:expprop} we added an important new exponential mirror solution with 
interesting properties. This leaves only one ``exponential'' unconsidered; in addition to 
exponential acceleration in $\tau$, $u$, $t$ and $x$, for completeness we now investigate 
the only clock not yet used for exponential acceleration: advanced time $v$ (not to be confused with 
velocity).  

For advanced time $v=t+z$, the proper acceleration behavior $\al(v)=\kp e^{\kp v}$ implies $\eta=\kappa v$. Since 
\be 
\frac{d\eta}{d\tau}=e^{\eta}\frac{d\eta}{dv}=\kp e^{\kp v} \ , 
\ee 
then $\kp\tau=-e^{-\eta}$ and we have the interesting property that the acceleration is scale independent, i.e.\ $\al(\tau)=-1/\tau$. Recall that in \cite{paper2} such scale independence -- but with a positive sign -- 
was shown to give eternal thermality of the radiation. 

This raises a second interesting aspect: from Eq.~A19 of \cite{paper2} we had obtained eternal 
thermality from 
$\al(v)=-(1/2)\sqrt{\kp/|v|}$. That is effectively the back side of the Carlitz-Willey 
eternally thermal moving mirror \cite{Carlitz:1986nh,Good:2013lca,Good:2012cp}. 
Both seem to be solutions, hinting at a potential ``duality'' (not in a strict 
mathematical sense) in 
the representation. Pursuing this further, Eq.~A18 of \cite{paper2} showed that exponential 
acceleration in a $u$ clock is also thermal $\al(\tau)\sim 1/\tau$. We have verified that 
$\al(u)=(1/2)\sqrt{\kp/|u|}$ gives the same $\al(\tau)=1/\tau$ solution as exponential in $u$. 
So $v\leftrightarrow e^{-v}$, and similar for $u$, seem to be related for these forms. 


Table~\ref{tab:expon} summarizes the complete family of mirrors with acceleration exponential 
in the various time variables. The expressions for energy flux have 
similarities to each other, with the exponential in $u$ having constant, thermal flux (Carlitz-Willey). 
The entropy flux, proportional to $\eta$, will in all these cases asymptotically diverge in contrast to the finite energy of the proper time exponentially accelerated mirror of Section \ref{sec:expprop}.

\renewcommand{\arraystretch}{1.5} 
\noindent
\centering
\begin{table*}[t]
\caption{Summary of acceleration, rapidity, and energy flux properties for the exponential acceleration family.} 
\label{tab:expon}
\begin{tabular}{|l||*{5}{c|}}\hline
\backslashbox[38mm]{Observable}{Coordinate}
&\makebox[3em]{$t$}&\makebox[3em]{$x$}&\makebox[3em]{$u$}
&\makebox[3em]{$v$}&\makebox[3em]{$\tau$}\\\hline\hline
Proper acceleration, $\alpha$ &$-\kappa e^{\kappa t}$ & $-\kappa e^{-\kappa x} $ &$ -\frac{\kappa}{2} e^{\frac{\kappa u}{2}} $& $\frac{\kappa}{2}e^{\frac{\kappa v}{2}} $ &$ -\kappa e^{\kappa \tau}$ \\\hline
Rapidity, $\eta$ & $\kappa x$ &$ -\kappa t$ &$-\frac{\kappa u}{2}  $ & $\frac{\kappa v}{2}$ & $-e^{\kappa \tau} $\\ \hline
\rule{0pt}{1.3\normalbaselineskip}Energy flux, $F$ & $ \frac{\kappa ^2}{48\pi}(1-e^{4\eta})$ &$\frac{\kappa^2}{48\pi}(1-e^{4 \eta})$&$\frac{\kappa^2}{48\pi} $&$ -\frac{\kappa^2}{48\pi}e^{4\eta} $&$-\frac{\kappa^2}{12\pi}\eta e^{2\eta}$\rule[-2.ex]{0pt}{1.3\normalbaselineskip}\\
\hline
\end{tabular}
\end{table*}


\end{document}